\documentclass[a4paper,12pt]{article}

\usepackage[latin2]{inputenc}
\usepackage{graphicx}
\usepackage{amstext}
\usepackage[]{ntheorem}
\usepackage{fancyhdr}
\usepackage{amsfonts}
\usepackage{amssymb}
\usepackage{amsmath,chemarrow}
\usepackage{array}
\usepackage{multirow}

\usepackage{t1enc}
\usepackage{tikz}
\usepackage{hyperref}
\usepackage[affil-it]{authblk}

\title{Strategic bidding via the interplay of minimum income condition orders in day-ahead power exchanges}

\author{D\'{a}vid Csercsik}

\affil{P\'{a}zm\'{a}ny P\'{e}ter Catholic University\\ Faculty of Information Technology and Bionics\\ Pr\'{a}ter~u. 50/A 1083 Budapest, Hungary \\
              Tel.: +36-1 886 47 00
              Fax: +36-1 886 47 24\\
              \emph{csercsik@itk.ppke.hu}}

\begin{document}

\maketitle

\abstract{In this paper we study the so-called minimum income condition order, which is used in some day-ahead electricity power exchanges to represent the production-related costs of generating units. This order belongs to the family of complex orders, which imply non-convexities in the market clearing problem. We demonstrate via simple numerical examples that if more of such bids are present in the market, their interplay may open the possibility of strategic bidding. More precisely, we show that by the manipulation of bid parameters, a strategic player may increase its own profit and potentially induce the deactivation of an other minimum income condition order, which would be accepted under truthful bidding. Furthermore, we show that if we modify the objective function used in the market clearing according to principles suggested in the literature, it is possible to prevent the possibility of such strategic bidding, but the modification raises other issues.}

\section*{Nomenclature}
\begin{table}[h!]
\begin{tabular}{|c|c|}
  \hline
  MCP & Market clearing price \\
  DAPX & Day-ahead power exchange \\
  TSW & Total social welfare \\
  MIC & Minimum income condition \\
  FT & Fixed term \\
  VT & Variable term \\
  \hline
\end{tabular}
\end{table}

\section{Introduction}
\label{sec_introduction}
If one investigates trading and pricing mechanisms in various electricity markets around the globe, it may be recognized that despite local market integration advancements and the convergence implied by them, the evolution of individual markets resulted in a diverse set of mechanisms and approaches \cite{oksanen2009electricity,sioshansi2011competitive,imran2014technical}. Moreover,
additional allocation and pricing mechanisms emerged in relation with electricity trade, as balancing markets \\ \cite{singh1999competitive} and transmission-related allocation and pricing \\ \cite{pan2000review}.
If we focus on short-term trading of electricity on competitive markets and its diversity among continents, we must recall the well-known paradigm that the general approach of such markets in the US is dominantly based on the unit commitment approach \cite{padhy2004unit}.  This paradigm basically means that generators submit their production characteristics to a independent system operator (ISO), who determines their schedule, according to consumption demands and to technological and economic constraints.
In contrast, in the EU, portfolio-bidding markets, or day-ahead power exchanges (DAPXs) are operating, where self-scheduling generating units and further market participants have the possibility to act as active market players and submit bids to two-sided markets, which are cleared in order to obtain zonal market clearing prices. In the general framework these markets are coupled, and the clearing mechanisms also take transmission constraints into account \cite{chatzigiannis2016european,biskas2013european_1,biskas2013european_2}. In this paper, we focus on the latter, European-type market structure.

\subsection{Simple hourly bids in DAPXs}
\label{subsec_hourly_bids}

The fundamental setting of these two-sided multiunit markets is very simple: Participants on the supply and the demand side  submit bids characterized by quantity ($q$) and price per unit ($p$) for the respective trading period(s) (typically hour) of the following day, which in general may be fully or also partially accepted, according to the resulting market clearing price (MCP) of each hour. In this basic setup, we practically look for the intersection point of demand and supply curves, which ensures the balance of consumption and production. This way, in each period maximum one bid is partially accepted, which determines the MCP.

\begin{figure}[h!]
  \centering
  \includegraphics[width=7cm]{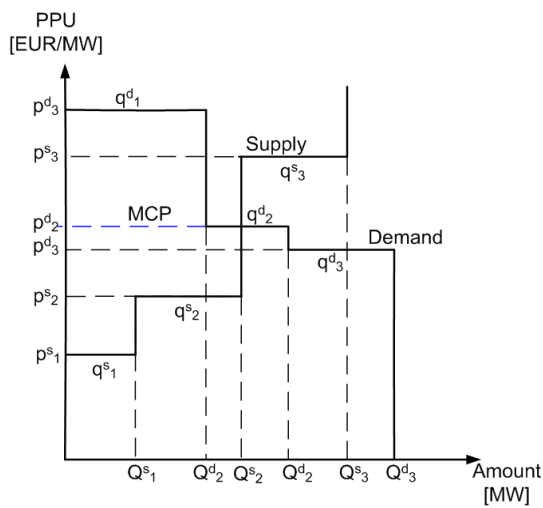}~
  \caption{Fundamental scheme of the day-ahead electricity spot market for a trading period. $Q^s_i$ and $Q^d_i$ stand for the cumulative quantities, i.e. $Q^s_i=\sum_{j=1}^i q^s_j$ and $Q^d_i=-\sum_{j=1}^i q^d_j$.}\label{Fig_basic_spot}
\end{figure}

Figure \ref{Fig_basic_spot} shows a simple example of such a market clearing for a single-period, single-zone case. We assume 3-3 bids on the supply and demand side denoted by S1-S3 and D1-D3 respectively.
The supply and demand curves depict bids sorted by their price (in increasing/decreasing order).

The bid quantity of bid $i$ of type $w$ ($w \in \{s,d\}$) is denoted by $q^w_i$, while its price is denoted by $p^w_i$. Demand bids are considered with negative quantity ($q^d_i<0$). Cumulative quantities, defining the breakpoints of the supply and demand curve are denoted by $Q^s_i$ and $Q^d_i$ respectively.

\begin{equation}
Q^s_i = \sum_{j=1}^i q^s_j~~~~~Q^d_i = -\sum_{j=1}^i q^d_j
\end{equation}

In this particular case, D2 (the second demand bid) is partially accepted, thus its bid price determines the MCP ($MCP=p^d_2$), and the traded quantity equals to $Q^s_2=q^s_1+q^s_2$.

We solved this simple case by determining the intersection point of the supply and the demand curves.
As we will see in the next subsection (\ref{subsec_CO}), special bids in these markets may be present, which do not allow partial acceptance. In this case, the curve-intersection approach may fail, if the intersection point hits one of these bids.

However, the intersection point on Figure \ref{Fig_basic_spot} has an other important interpretation.
Let us define acceptance values for each bid: $y^w_i \in [0,1]$ denotes the acceptance ratio of bid $i$ of type $w$ ($w \in \{s,d\}$).
The acceptance values which determine the intersection point in this case are
\begin{equation}
\label{sol_example_1}
y^s_1=1~~~y^s_2=1~~~y^s_3=0~~~y^d_1=1~~~y^d_2=\frac{3}{7}~~~y^d_3=0~.
\end{equation}

In electricity trade, the supply must always meet the demand, thus we require the balance
\begin{equation} \label{balance}
\sum_i y^s_i q^s_i + \sum_i y^d_i q^d_i =0~~~.
\end{equation}

 Let us furthermore define the concept of the total social welfare (TSW). TSW in this context is interpreted as the total utility of consumption, minus the total cost of production, formally
\begin{equation}
\label{TSW_basic}
 - \sum_i y^d_i q^d_i p^d_i - \sum_i y^s_i q^s_i p^s_i ~~~.
\end{equation}
(Remember that $q^d_i<0$.)

This quantity equals to the area between the accepted part of the demand and supply curves.
It is easy to see that if we start from the point defined by the values (\ref{sol_example_1}) and decrease or increase the total traded quantity (the supply-demand balance must still hold), the TSW is strictly decreased. In fact, the values described in (\ref{sol_example_1}) are exactly those values, which maximize the total social welfare of the market, assuming the balance constraint \ref{balance}.

This concept of TSW maximization may be generalized to more complex cases. Indeed, this principle is
usually a fundamental element of European-type market clearing mechanisms.

In general, if multiple trading periods are considered and no interdependencies arise between the periods, the above approach may be applied for each of the periods independently. In this paper, we will assume a simple two-period case, but as we will see in the following, the characteristic order types of electricity trade will define interdependencies between the periods.

\subsection{Complex orders in electricity markets}
\label{subsec_CO}

It is easy to see that the setup detailed above does not consider technological constraints of generating units: It is possible that the resulting MCPs imply that a unit submitting bids for two consecutive hours must produce at full capacity in the first hour and shut down in the second (if its bid is fully accepted in the first period, and fully rejected in the second), while the technological constraints of the unit make this impossible. The first approach to address these problems has been the introduction of so-called block orders \cite{meeus2009block}, which connect multiple bids submitted for various periods and they must be fully accepted or rejected in all of the respective periods (in other words, they are characterized by the '\emph{fill-or-kill}' condition).
These bids imply non-convexities (integer variables) in the market clearing problem \cite{madani2016non}, making the efficient clearing of large scale markets challenging \cite{madani2018revisiting}.
To guarantee the existence of MCP when block orders are allowed, we must allow their deactivation, regardless of their bid prices \cite{madani2017revisiting}. This may result in so-called paradoxically rejected block orders \cite{madani2014minimizing}, the rejection of which seemingly contradicts to the resulting market clearing prices (but in fact, they can not be accepted without the violation of other constraints, since if they get accepted the MCP will not exist anymore \cite{madani2016non}).

Block orders are also beneficial for incorporating the non-negligible start-up cost of the generating units. The simple cost model of generating units usually includes a fixed term (FT) corresponding to start-up costs and a variable term (VT), which is interpreted as the linear coefficient describing the connection between the generated quantity and the variable (fuel) cost of the units (see e.g. \cite{richstein2018auction}).
The efficient implementation and market effects of block orders have been discussed in the literature
\cite{meeus2009block,madani2014minimizing,madani2015computationally}.

\subsubsection{Minimum income condition orders}

In addition to block orders, one may find further so-called complex orders and products in today's practical electricity market implementations \cite{sleisz2016complex,dourbois2015european,van2011linear,chatzigiannis2016european_IEEE}. One of these complex orders is the so-called \emph{Minimum Income Condition (MIC)} order.
The MIC was first introduced in the Spanish electricity market
\cite{Contreras2001} and since then it became quite commonly used in various market models
\cite{GARCIABERTRAND2006457,Lam018european,dourbois2015european,dourbois2017novel,operator2013daily,NL2016}, including
novel approaches which aim to provide optimization-based framework for the optimal joint energy and reserves market clearing \cite{koltsaklis2018incorporating}.

 Nevertheless, the
necessity \cite{poli2011clearing}, the effects on market outcomes \cite{ruiz2012pricing,madani2016non,gil2017minimum} and the efficient implementation of MIC orders \cite{polgari2015new,sleisz2015challenges,sleisz2015efficient,madani2018revisiting,sleisz2019new} are still subject to ongoing debates and studies.

EUPHEMIA, the market-coupling tool which was brought to life by European market integration trends, and serves as a kind of reference for European market design approaches,
also includes MIC orders. In its public description \cite{euphemia2015}, Minimum Income Condition (MIC) orders are defined
as supply orders consisting of several hourly step bids (elementary bids) for
potentially different market hours, which are connected by the MIC which prescribes that the overall income of the MIC order must cover its given costs. These costs are defined by a fix term (representing the startup cost of a power plant) and a variable term multiplied by the total assigned production volume (representing the operation cost per MWh of a power plant).
Formally, the Minimum Income Condition constraint is defined by two parameters:
\begin{itemize}
  \item A fix term (FT) in Euros
  \item A variable term (VT) in Euros per accepted MWh.
\end{itemize}

In the final solution, MIC orders may be activated or deactivated (as a whole):
\begin{itemize}
  \item In case a MIC order is activated, each of the hourly sub-orders of the MIC behaves like any other hourly order, which means that they are accepted if and only if the MCP is higher or equal to the bid price.
  \item In case a MIC order is deactivated, each of the hourly sub-orders of the MIC is fully rejected, even the MCP is higher or equal to the bid price.
\end{itemize}

We can see that the MIC condition links multiple hourly bids, and the necessary condition for the acceptance of these bids is the activation of the MIC order, which can be described by a binary variable.

\subsection{Incentive-compatibility and its relevance in electricity-related markets}

The concept of incentive-compatibility is originating from the 70's \cite{hurwicz1973design,groves1987incentive}, and it is related to the evaluation of allocation mechanisms under the assumption of strategic behavior of participants.
The topic is discussed in auction theory \cite{klemperer2004auctions}, however most of the results in
this field correspond to single-unit auctions of indivisible goods \cite{roth1982incentive}, while in the case of electricity markets a multi-unit auction framework applies. Let us however note that the problem of simultaneous allocation of multiple indivisible goods with complementarities is addressed in the framework of combinatorial auctions \cite{de2003combinatorial,cramton2007overview}.

As formulated by \cite{nisan2007}, '\emph{A mechanism is called incentive-compatible if every participant can achieve the best outcome to themselves just by acting according to their true preferences}'.
As the original problem statement assumes indivisible goods, which does not hold in multi-unit electricity auctions, preferences translate to evaluations in our case: Considering the simple example depicted in Fig. \ref{Fig_basic_spot}, in ideal case, bidders on the demand side bid their real consumption utilities and bidders on the supply side bid their real marginal costs.
We consider strategic bidding compared to this reference case of truthful bidding.

\begin{figure}[h!]
  \centering
  \includegraphics[width=7cm]{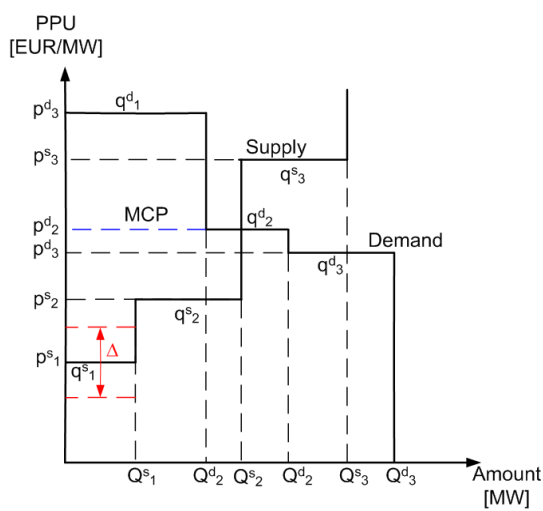}~
  \caption{Possible small variations of the bid price of supply bid 1 ($\Delta$) has no effect on the MCP, thus the acceptance and payoff supply bid 1 is not affected.}\label{Fig_basic_spot_v2}
\end{figure}

Figure \ref{Fig_basic_spot_v2} demonstrates, why the standard, marginal (i.e. MCP based) clearing model of multi-unit electricity markets is considered to be \emph{practically} incentive-compatible.

If we perturb the bid price of supply bid 1 (S1) by $\Delta$ as depicted in the picture, the outcome of the auction does not change:
The MCP determining the set of accepted/rejected bids will be the same as before. In addition, as bids are paid off according to the MCP, the resulting utility of the deviating player also remains the same. If e.g. the bid price is increased, the nominal income of the bid is increased, but the surplus resulting from the difference of the MCP and the bid price is decreased by the same amount.

The above deduction is on the one hand only true for small deviations (e.g. in our case we assume that for the modified value $\hat{p}^s_1$ the inequality $\hat{p}^s_1<p^s_2$ still holds) and on the other hand it is not true for all bids. As in the proposed case the second demand bid (D2) sets the MCP, $p^d_2$ is in fact affecting the market clearing price, and if the bidder of D2 decreases $p^d_2$, it can effectively increase its own surplus (as long as the inequality $p^d_2 \geq p^d_3$ holds).

If we consider an other bid than D2, but perturb the price so much that it changes the ordering of bids, and the actual bid will be exactly the price-setter (e.g. if we decrease $p^d_1$ to $p^d_3<\hat{p}^d_1<p^d_2$), the same effect arises.

In practice, the number of standard bids for any period in DAPXs is high (several thousand or tens of thousands), and as in the bid submission process the actual other bids are not known, at first glance it seems unlikely that such manipulation can be effectively carried out.
On the other hand, as recent examples have shown \cite{Moylan2014}, the number of big players may be limited, and high proportion of bids may originate from the same players. In this case, oligopolistic behavior and related phenomena may emerge on electricity markets
 \cite{david2001market}.

Regarding MIC orders, according to the publicly available data of OMIE (the Spanish DAPX), where this formulation is used (\url{https://www.omie.es}), the number of these orders is significantly lower, about 75-85 per day.

\begin{figure}[h!]
  \centering
  \includegraphics[width=7cm]{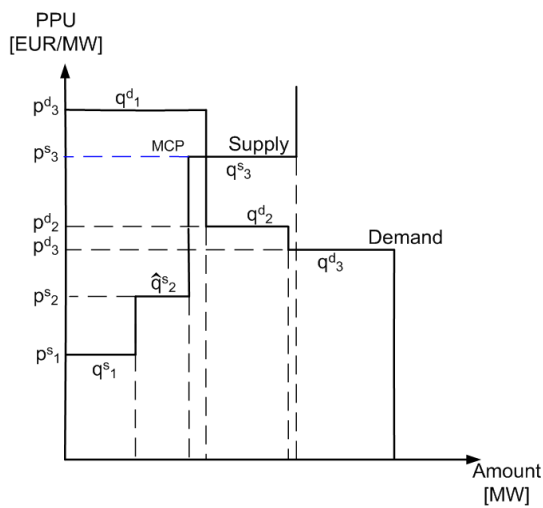}~
  \caption{Reduction of supply: The quantity of the bid S2 is modified to the half of its original value. This causes the MCP to increase.}\label{Fig_basic_spot_v3}
\end{figure}

In the above reasoning, we assumed the manipulation of bid price. This is however not the only alternative. Let us assume a market scenario, where there are some large suppliers with significant bid quantities, like S2 in Fig. \ref{Fig_basic_spot}). As depicted in Fig. \ref{Fig_basic_spot_v3}, if $q^s_2$ is reduced to $\hat{q}^s_2=0.5~q^s_2$, the intersection point of the two curves will change, the MCP is increased, and the bidder of S2 receives significantly more payoff for one unit of energy.
This phenomena is also termed as \emph{capacity withholding}, discussed by \cite{esmaeili2016analysis} in a similar auction-based but also network-constrained framework.
The possibility also arises on the other side (demand reduction), although in electricity markets, large producers (i.e. big power plants) are more prevalent than large consumers.

While in this very case the supply is reduced in the market, in order to get a higher payoff, similar effects may arise if multiple participants are competing for a fixed set of goods, as in the case of
treasury and spectrum auctions. In this case, as discussed by \cite{ausubel2002demand}, we talk about demand reduction.

Up to this point we discussed cases in which one participant unilaterally changes its bidding behavior to improve its payoff, while it was assumed that the rest of the bids is unchanged.
The question however may be formulated in a more general context as well, where all submitted bids are subject to strategic behavior. The paper of \cite{aliabadi2016determining} considers power generation companies (GenCos) located in a network as leaders of a Stackelberg type game, in which the independent system operator (ISO) plays the role of the follower. The paper develops a bi-level mathematical programming framework to model the market clearing mechanism of the ISO where the behavior of GenCos and network constraints are taken into account. In this model, the authors are able to describe the collusive behavior of GenCos, and also provide numerical examples demonstrating the possible uniqueness/non uniqueness of Nash Equilibria.
As we will see later, our main aim in this paper is to show how the minimum income condition orders may be used for strategic bidding. Potential equilibrium problems resulting from such strategic behavior are not the main focus of this study, they are discussed only marginally.

\subsubsection{Pay-as-bid auction and incentive-compatibility}

Let us note that in addition to the marginal clearing model described in subsection
\ref{subsec_hourly_bids}, the pay-as-bid method is also applies in the case of some electricity related markets, like for example the Iranian electricity market \\ (\url{http://www.irema.ir/trading/day-ahead-market}).
 However, as discussed by \cite{tierney2008uniform}, in this case, in contrast to the marginal clearing market model, all participants do have clear incentives: Suppliers aim to raise their bid prices up to the maximum acceptable level to earn the most payoff. This results in a flattened supply curve, and according to \cite{tierney2008uniform}, it exacerbates market competitiveness.

In other studies, agent based simulations were used to determine optimal strategic bidding behavior and market efficiency in the context of pay as bid vs marginal pricing \cite{xiong2004multi,bakirtzis2006agent,bower2001experimental,liu2012multi,aliabadi2017agent}.

\subsection{Contribution and structure of the paper}
The possibilities of strategic bidding via MIC orders have not been explicitly discussed in the literature.
In this paper, our aim is to show that under the current practice, bidders submitting MIC bids may have additional incentives for strategic bidding, if multiple such bids are present on the actual market.
To demonstrate this, based on the widely used EUPHEMIA framework and the respective definition and formalism of MIC orders, we introduce a simple market clearing framework for a two period market, where the phenomenon may be studied in its purest form.

In section \ref{sec_MM}, we introduce the computational implementation of the market in detail. Section
\ref{sec_results} demonstrates how strategic bidding may increase the payoff of MIC bids via the interplay of such bids, and also analyses the scenario, when a modification in the objective function of the market clearing algorithm is introduced to address this issue. Section \ref{sec_discussion} evaluates and discusses the proposed results, and finally section \ref{sec_conclusions} concludes and drafts future prospects of the work.

\section{Materials and Methods}
\label{sec_MM}

In the next subsection, we introduce the market model in which the interplay of MIC orders is studied.
To clarify our terminology, MIC \emph{orders} are composed of multiple hourly \emph{bids} (termed 'sub-orders' in the EUPHEMIA description cited in subsection \ref{subsec_CO}), belonging to different trading periods (hours).

\subsection{Computational implementation}
\label{comp_impl}

The market clearing in DAPXs is implemented as an optimization problem. In this section, we introduce the components (variables) of this problem and formulate the corresponding constraints and the objective function.

We assume a simplified single-zone market model, where only two time periods are considered. In addition, we assume that only two types of bids are present on the market:
\begin{itemize}
\item \textbf{Simple hourly bids}. The acceptance of these bids is solely determined by the MCP of the respective period.
\item \textbf{Hourly bids belonging to complex MIC orders}. The acceptance of these bids depend not only on the MCP of the period to which the bid belongs, but on the total income of the order, which, in turn, depends on the MCP values of other periods as well.
\end{itemize}

Since MIC conditions of complex orders define interdependencies between trading periods, the bids submitted for various periods must be cleared simultaneously.

The computational form of the market model used in this paper includes the following variables:
\begin{itemize}
  \item Market clearing prices (MCPs) of the two trading periods, denoted by $MCP_1$ and $MCP_2$ respectively. In the current paper we assume that every $MCP$ is nonnegative.
  \item Acceptance variables of simple hourly supply bids. The acceptance variable of the i-th simple supply bid is denoted by $y^s_i$. All acceptance variables are bounded as $0 \leq y \leq 1$.
  \item Acceptance variables of simple hourly demand bids. The acceptance variable of the i-th simple demand bid is denoted by $y^d_i$.
  \item Acceptance variables of hourly bids belonging to complex orders. The acceptance variable corresponding to the i-th component of complex order $c$ is denoted by $y^c_i$
  \item Variables corresponding to the income of individual bids of complex orders. The income of bid $y^c_i$ is denoted by $I^c_i$.
  \item Auxiliary integer variables corresponding to the big-M implementations of logical implications.
  The vector of these variables is denoted by $z$.
  \end{itemize}

\subsubsection{Simple hourly orders}
\label{subsec_Simple hourly orders}
The acceptance constraints in the case of simple hourly supply bids may be written as
\begin{align}
& y^s_i > 0 ~~~\rightarrow~~~ p^s_i \leq MCP_t \nonumber \\
&y^s_i < 1 ~~~\rightarrow~~~ MCP_t \leq p^s_i \label{BA_simple_suppy}
\end{align}
where $p^s_i$ is the bid price of the simple hourly supply bid $i$, and $MCP_t$ denotes the $MCP$
of period $t$, for which the bid is submitted ($t \in \{1,2 \}$).

The implications may be easily included in the MILP formulation. Let us consider e.g. the first implication of eq. \ref{BA_simple_suppy}, which is equivalent to
\begin{equation}\label{impl_in_MILP_1}
p^s_i \leq MCP_t~~~or~~~y^s_i \leq 0~~.
\end{equation}
The equivalent of the logical expression (\ref{impl_in_MILP_1}) is the set of the inequalities (\ref{impl_in_MILP_2}), where
$z\in\{0,1 \}$ is an auxiliary integer variable and $\overline{MCP}$ is the upper bound of the variable $MCP$.
\begin{align}
&p^s_i - z \overline{MCP}_t \leq MCP_t \nonumber \\
&y^s_i - (1-z) \leq 0 \label{impl_in_MILP_2}
\end{align}
We can use the variable $z$ to 'cancel' one of the inequalities of \ref{impl_in_MILP_1}, but not both of them. In the following we assume that all implications are implemented using the above 'bigM' method (where the bigM refers to $\overline{MCP}$ the upper bound of $MCP$ used in the formulation).

Similarly to supply bids, in the case of simple hourly demand bids, the constraints may be written as
\begin{align}
&y^d_i > 0 ~~~\rightarrow~~~  MCP_t \leq p^d_i\nonumber\\
&y^d_i < 1 ~~~\rightarrow~~~ p^d_i \leq MCP_t, \label{BA_simple_demand}
\end{align}
where $p^S_i$ is the bid price of the simple hourly demand bid $i$.



\subsubsection{Bids of complex orders}
\label{subsec_Bids of complex orders}
The first part of the constraints described in the formula (\ref{BA_simple_suppy}) is also active in the case of supply bids belonging to complex orders:

\begin{equation}\label{BA_complex}
y^c_i > 0 ~~~\rightarrow~~~ p^c_i \leq MCP_t,
\end{equation}
where $p^c_i$ is the bid price of the i-th component of the complex order $c$, corresponding to time period $t$.

The considerations of an MIC bid described in subsection \ref{subsec_CO} are formulated in the optimization framework of market-clearing algorithms as

\begin{align}\label{MIC_cond_basic_accept}
  \sum_{i} y^c_i >0 \rightarrow FT_c~~+~~VT_c \sum_{i} q^c_i y^c_i \leq I^c
\end{align}
where $y_k$ is the acceptance indicator of the elementary bid $k$ belonging to set of bids of the complex order $c$. $FT_c$ and $VT_c$ denote the fixed and variable cost terms of complex order $c$.
$q^c_i$ is the bid quantity of bid $y^c_i$, and $I^c$ is the variable representing the total income of the complex (MIC) order $c$, which may be calculated as

\begin{equation}\label{total_income_of_MIC_order}
  I^c= \sum_i I^c_i
\end{equation}

Intuitively $I^c_i$ may be calculated as

\begin{align}
I^c_i = MCP_t~ q^c_i~ y^c_i \label{income_naive}
\end{align}
where $q^c_i$ stands for the quantity of the bid $y^c_i$.

Equation (\ref{income_naive}) however includes a quadratic expression of variables, namely the product of
$MCP_t$ and $y^c_i$, the implementation of which which would result in a computationally demanding quadratically constrained problem (MIQCP). To overcome this issue, and obtain a linear form of expressions, following \cite{sleisz2015efficient,sleisz2019new}, we formulate the expressions for income as

\begin{align}
&y^c_i >0 ~~\rightarrow ~~ I^c_i = y^c_i q^c_i p^c_i + q^c_i MCP_t - q^c_i p^c_i \label{income_formulation_1}\\
&y^c_i <1 ~~\rightarrow ~~ I^c_i = y^c_i q^c_i p^c_i \label{income_formulation_2}
\end{align}
%
As described by \cite{sleisz2015efficient}, taking into account the bid acceptance rule described in (\ref{BA_complex}), three possibilities may arise:
\begin{enumerate}
  \item If the bid is entirely accepted ($y^c_i=1$), $I^c_i$ equals the product of $q^c_i$ and $MCP_t$ according to (\ref{income_formulation_1}).
  \item If the bid is partially accepted ($MCP_t=p^c_i$), $I^c_i$ equals to $y^c_i q^c_i p^c_i$. Both (\ref{income_formulation_1}) and (\ref{income_formulation_2}) are active in this case and they result in the same inequality.
  \item And finally, if the bid is entirely rejected ($y^c_i=0$), according to (\ref{income_formulation_2}) $I^c_i=0$.
\end{enumerate}

\subsubsection{Power balance}
\label{subsec_Power_balance}
Formula (\ref{eq_power_balance}) describes that the quantity of accepted supply bids must be equal
the quantity of accepted demand bids for all periods. Let us note again that the quantity of demand bids is negative by definition.

\begin{equation}\label{eq_power_balance}
  \sum_{i \in B_t} y^s_i + \sum_{\{c,i\} \in B_t} y^c_i + \sum_{i \in B_t} y^d_i  = 0~~~~~\forall t
\end{equation}
where $B_t$ denotes the set of bids corresponding to period $t$.

\subsubsection{The objective function}
\label{subsec_nominal_TSW}

Following the fundamental concepts of day-ahead electricity auctions \cite{madani2017revisiting}, the objective function of the problem is to maximize the total social welfare (TSW), defined as

\begin{equation}\label{TSW}
  TSW = - \sum_i y^d_i q^d_i p^d_i - \sum_i y^s_i q^s_i p^s_i - \sum_{c,i} y^c_i q^c_i p^c_i
  \end{equation}

In other words, the TSW is the total utility of consumption minus the total cost of production, in the context of the acceptance/rejection of hourly bids. As $q^D_i<0$ by definition for all $i$, the corresponding terms must be multiplied with -1.

Let us note that this objective is in accordance with the concept used in EUPHEMIA \cite{euphemia2015}, where a somewhat different terminology is used. The EUPHEMIA description aims to maximize the sum of the \emph{consumer surplus} and the \emph{producer surplus}, which is in fact the TSW. The consumer surplus ($CS$) and the producer surplus ($PS$) may be derived as

 \begin{align}
  CS = \sum_t \sum_{i \in B_t} y^d_i q^d_i (p^d_i-MCP_t) \nonumber \\
  PS = \sum_t \sum_{i \in B_t} y^s_i q^s_i (MCP_t-p^s_i) \label{surplusses}
  \end{align}

If one considers a simple case without block bids, as depicted Fig. \ref{Fig_basic_spot}, the TSW is the area between the demand and the supply curve, considering those parts which are leftmost of the intersection point. The MCP divides this area into two parts: The upper is CS, while the lower is PS ($TSW=CS+PS$).

Overall, the market clearing problem may be summarized as

 \begin{align}
&   \max_{MCP,Y^S_i,Y^D_i,Y^c_i, Z} TSW~~~~\text{wrt.} \nonumber\\
 &  \sum_{i \in B_t} y^s_i + \sum_{\{c,i\} \in B_t} y^c_i + \sum_{i \in B_t} y^d_i  = 0~~~~~\forall t \nonumber\\
 & y^s_i > 0 ~~~\rightarrow~~~ p^s_i \leq MCP_t ~~\forall i \nonumber \\
&y^s_i < 1 ~~~\rightarrow~~~ MCP_t \leq p^s_i ~~\forall i \nonumber\\
&y^d_i > 0 ~~~\rightarrow~~~  MCP_t \leq p^d_i~~\forall i \nonumber\\
&y^d_i < 1 ~~~\rightarrow~~~ p^d_i \leq MCP_t ~~\forall i \nonumber\\
&y^c_i > 0 ~~~\rightarrow~~~ p^c_i \leq MCP_t ~~\forall i  \nonumber\\
&\sum_{i} y^c_i >0 \rightarrow FT_c~~+~~VT_c \sum_{i} q^c_i y^c_i \leq I^c ~~\forall c  \nonumber\\
&y^c_i >0 ~~\rightarrow ~~ I^c_i = y^c_i q^c_i p^c_i + q^c_i MCP_t - q^c_i p^c_i ~~\forall c,~\forall i \nonumber\\
&y^c_i <1 ~~\rightarrow ~~ I^c_i = y^c_i q^c_i p^c_i ~~\forall c,~\forall i \nonumber\\
  \label{optimization_problem}
 \end{align}

 where MCP is the vector of market clearing prices for the two periods, $MCP_t$ denotes the MCP for the respective time period (for which the bid is submitted). $Y^S_i$ and $Y^D_i$ are the vectors of acceptance variables corresponding to simple supply and demand bids, $Y^c_i$ is the vector of bids belonging to complex orders, $I$ is the vector of incomes of these bids (auxiliary variables), and $Z$ holds the auxiliary binary variables used in the big-M formulations of the logical implications (\ref{BA_simple_suppy},\ref{BA_simple_demand},\ref{BA_complex},\ref{MIC_cond_basic_accept},\ref{income_formulation_1} and \ref{income_formulation_2}.

\section{Results}
\label{sec_results}
In this section, we introduce a simple example bid set in a 2-period example, to demonstrate that due to the special interplay between MIC bids, participants may have incentives to bid false production cost.
Let us assume the bids described in Table \ref{Table_bids_1}.

\begin{table}[h!]
\begin{center}
\begin{tabular}{|c|c|c|c|c|}
  \hline
  ID & t &q & p & var\\ \hline
  S1 & 1 & 2 & 5 & $y^s_1$\\
  S2 & 1 & 2 & 6 & $y^s_2$\\
  S3 & 2 & 2 & 5 & $y^s_3$\\
  S4 & 2 & 2 & 6 & $y^s_4$\\
  S5 & 1 & 2 & 1 & $y^{c1}_1$\\
  S6 & 2 & 2 & 1 & $y^{c1}_2$\\
  S7 & 1 & 2 & 4 & $y^{c2}_1$\\
  S8 & 2 & 2 & 4 & $y^{c2}_2$\\
  D1 & 1 & -5 & 10 & $y^{d}_1$\\
  D2 & 2 & -5 & 10 & $y^{d}_2$\\
  \hline
\end{tabular}
\end{center}
\caption{Hourly bids of example I: Parameters and corresponding variables. \label{Table_bids_1}}
\end{table}
We suppose that S5-S6 and S7-S8 are part of complex MIC orders ($c1$ and $c2$ respectively), while the rest of the bids are standard bids, cleared purely according to the resulting MCP.

Let us furthermore assume that the production cost parameters of the units corresponding to
$c1$ and $c2$ are described by the following parameters:
\begin{align}\label{FT_VT_nominal}
& FT_1=10~~~VT_1=2 \nonumber \\
&  FT_2=10~~~VT_2=2~~~.
\end{align}

We assume that the participant submitting the complex order $c1$ is the only strategic player, and we call this participant player 1 in the following.

\subsection{Case 1}
\label{Case_1}
In this case, we assume that player 1 submits its real production costs as parameters of the
complex order (i.e. submits the real $FT_1$ and $VT_1$ values, as described in (\ref{FT_VT_nominal})). As player 2 is not considered as a strategic player in this example, in the following we assume that $FT_2$ and $VT_2$ is always equal to the values in (\ref{FT_VT_nominal}). This case serves as the reference describing truthful bidding.

In this case, if we maximize the total social welfare based on the hourly bids (as described in subsection \ref{comp_impl} and as it is usual in the case of European portfolio-bidding type markets), we get the following result.
The MCP is 5 in both periods, and regarding the standard bids, the acceptance indicators are as
\begin{align}
\left(
  \begin{array}{c}
    y^s_{1} \\
    y^s_{2} \\
    y^s_{3} \\
    y^s_{4} \\
    y^d_{1} \\
    y^d_{2} \\
  \end{array}
\right)=
\left(
  \begin{array}{c}
    0.5 \\
    0 \\
    0.5 \\
    0 \\
    1 \\
    1 \\
  \end{array}
\right)~~~.
\end{align}

The acceptance values of the bids corresponding to complex orders $c1$ and $c2$ are

\begin{align}
\left(
  \begin{array}{c}
    y^{c1}_1 \\
    y^{c1}_2 \\
  \end{array}
\right)=
\left(
  \begin{array}{c}
    1 \\
    1 \\
  \end{array}
\right)
~~~~
\left(
  \begin{array}{c}
    y^{c2}_1 \\
    y^{c2}_1 \\
  \end{array}
\right)=
\left(
  \begin{array}{c}
    1 \\
    1 \\
  \end{array}
\right)
\end{align}

\begin{figure}[h!]
  \centering
  \includegraphics[width=10cm]{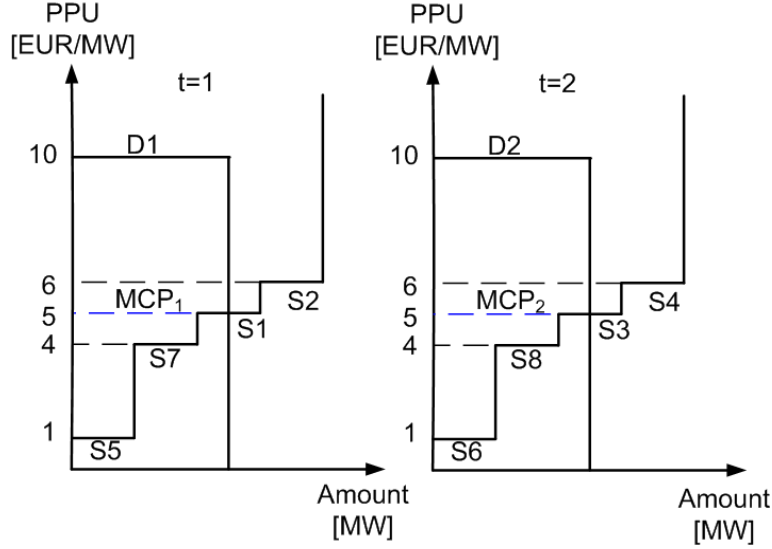}~
  \caption{Resulting dispatch in case 1.}\label{MIC_case_1}
\end{figure}

The total cost of MIC bid $c1$ is 18, while its total income is 20, thus the MIC condition holds.
The profit of the strategic player is 2 units in this case.
Similarly, the total cost of MIC bid $c2$ is 18, while its total income is also 20, thus the MIC condition holds here as well. The resulting dispatch is depicted in Fig. \ref{MIC_case_1}.

\subsection{Case 2}
\label{Case_2}
In this case, we assume that the parameters of the hourly bids are the same (as described in Table \ref{Table_bids_1}), while player 1 increases the submitted FT value from 10 to 14.
\begin{align}\label{MIC_ex_1b}
& FT_1=\textbf{14}~~~VT_1=2 \nonumber \\
&  FT_2=10~~~VT_2=2
\end{align}

In this case, it is impossible to accept both $c1$ and $c2$. If we consider the dispatch depicted in Fig. \ref{MIC_case_1}, which corresponds to the simultaneous acceptance of $c1$ and $c2$, we can see that the total income of $c1$ is still 20, but the respective total cost
(according to the reported parameters) is 22 (14 units from the fixed cost and 8 units from the variable cost), thus the market clearing algorithm will not allow this outcome.

If we perform the optimization of the market clearing in this case, we get the following results.
The MCP is 6 in both periods, and regarding the standard bids, we get the acceptance indicators
\begin{align}
\left(
  \begin{array}{c}
    y^s_{1} \\
    y^s_{2} \\
    y^s_{3} \\
    y^s_{4} \\
    y^d_{1} \\
    y^d_{2} \\
  \end{array}
\right)=
\left(
  \begin{array}{c}
    1 \\
    0.5 \\
    1 \\
    0.5 \\
    1 \\
    1 \\
  \end{array}
\right)~~,
\end{align}

while the values of the complex MIC bids $c1$ and $c2$ are

\begin{align}
\left(
  \begin{array}{c}
    y^{c1}_1 \\
    y^{c1}_2 \\
  \end{array}
\right)=
\left(
  \begin{array}{c}
    1 \\
    1 \\
  \end{array}
\right)
~~~~
\left(
  \begin{array}{c}
    y^{c2}_1 \\
    y^{c2}_2 \\
  \end{array}
\right)=
\left(
  \begin{array}{c}
    0 \\
    0 \\
  \end{array}
\right)
\end{align}

According to the submitted parameters, the total cost of MIC bid $c1$ is 22, while its total income is 24, thus the MIC condition holds.
Considering the real parameters, the production cost of $c1$ is still 18, thus the real profit of player 1 is increased from 2 to 6 compared to the truthful bidding case described in subsection \ref{Case_1}.
As no hourly bid of MIC order $c2$ is accepted, its cost is zero, thus the corresponding MIC holds.

\begin{figure}[h!]
  \centering
  \includegraphics[width=10cm]{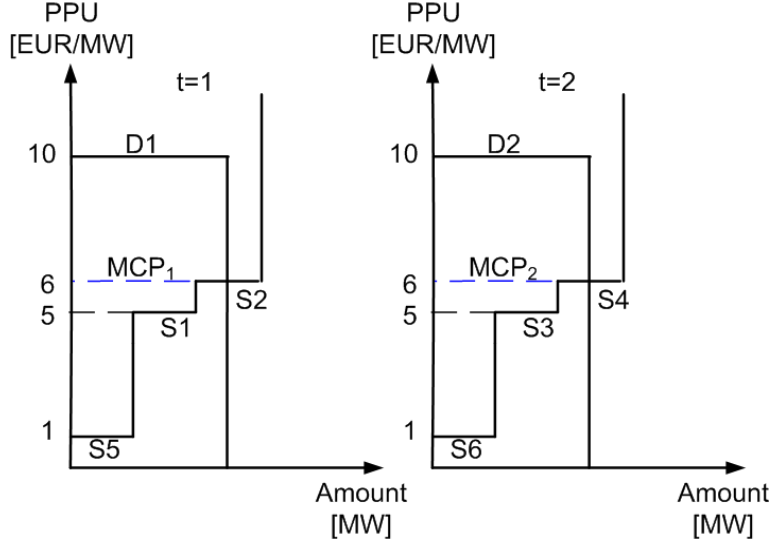}~
  \caption{Resulting dispatch in case 2.}\label{MIC_case_2}
\end{figure}

The resulting dispatch is depicted in Fig. \ref{MIC_case_2} As the complex order $c2$ is deactivated, its
hourly bids (S7 and S8) are not included in the supply curve.

We can see that by increasing the $FT$ value, player 1 has 'pushed out' MIC order S2, and increased its own profit. The reason behind this is the following:
The increase of $FT_1$ would intuitively imply the rejection of $c1$, because its MIC condition is not valid anymore. As the objective function (the TSW) is however determined on the basis of the hourly bids, the solver does not want to 'loose' the hourly bids S5 and S6 corresponding to $c1$, since, due to their low bid price, they significantly contribute to the TSW. A more optimal solution is to drop the bids $S7$ and $S8$ of $c2$. By the deactivation of the complex order $c2$, all corresponding constraints are fulfilled, and $c1$ may remain active.

In this case, $c2$ is a paradoxically rejected MIC order in this case. If we consider only the resulting MCPs, it seems that the hourly bids belonging to $c2$ may be (fully) accepted, and its MIC may be fulfilled. This is however not a feasible scenario, exactly as in the case of paradoxically rejected block orders \cite{madani2017revisiting}.

Let us note furthermore that any submitted FT value for $c1$ satisfying $12<FT_1 \leq 16$ will do the job for player 1, in the sense that in any such case the hourly bids of $c1$ will fully be accepted, while the order $c2$ will be deactivated. The resulting MCP and thus the (real) profit of player 1 is independent of the exact value of $FT_1$ in this case (if $12<FT_1 \leq 16$). If $FT_1 \leq 12$, the MIC condition will also be satisfied with the MCP of 5, thus the acceptance of $c2$ is allowed, and if $16 < FT_1$, the MIC of $c1$ will not hold, thus it will be deactivated and $c2$ will be accepted.

\subsection{Modifying the objective function}
\label{subsec_modified_TSW}

As the anomaly discussed above originates from the fact that the objective function is determined solely by the acceptance values and parameters of the hourly bids, a quite straightforward approach to resolve such problem is the modification of the objective function.
 In this case, the hourly bids corresponding to complex orders are not considered in the objective function. Instead of them, the cost of production in the case of complex orders is considered based on the given price parameters corresponding to fixed and variable terms (FT and VT). There are examples present in the literature, which follow a similar approach.
Start-up costs are included in the objective function in \cite{gabriel2013solving,ruiz2012pricing}, and \cite{madani2018revisiting} also consider the start-up costs in the objective, however the formulation is different in this case.

Formally, we can write the modified objective function as described in eq. (\ref{TSW_2}).

\begin{equation}\label{TSW_2}
  TSW = - \sum_i y^d_i q^d_i p^d_i - \sum_i y^s_i q^s_i p^s_i - \sum_c y^c FT_c - \sum_{c,i} y^c_i q^c_i VT_c
  \end{equation}
where $y^c\in \{0,1\}$ equals to 1 if the complex order $c$ is activated, $FT_c$ and $VT_c$ are the respectively the fixed and variable costs of complex bid $c$.

In the following, we will show that this modification of the objective function resolves the possibility of strategic bidding described in subsection \ref{Case_2}, but it implies a different kind of issue.

\subsubsection{Outcome of case 1 assuming the modified TSW}
\label{Case_1_mod}

If we apply the modified TSW described in eq. (\ref{TSW_2}) for the bid set described in subsection \ref{Case_1}, the results do not change. All the resulting acceptance indicators, MCPs and payoffs are the same as in the original case.

\subsubsection{Outcome of case 2 assuming the modified TSW}

On the other hand, if the modified TSW described in eq. (\ref{TSW_2}) is applied for the bid set described in subsection \ref{Case_2}, the results are affected. In this case, we get the following results.
The MCP is 6 in both periods, regarding the standard bids,
\begin{align}
\left(
  \begin{array}{c}
    y^s_{1} \\
    y^s_{2} \\
    y^s_{3} \\
    y^s_{4} \\
    y^d_{1} \\
    y^d_{2} \\
  \end{array}
\right)=
\left(
  \begin{array}{c}
    1 \\
    0.5 \\
    1 \\
    0.5 \\
    1 \\
    1 \\
  \end{array}
\right)
\end{align}

while the values of the complex MIC bids $c1$ and $c2$ are

\begin{align}
\left(
  \begin{array}{c}
    y^{c1}_1 \\
    y^{c1}_2 \\
  \end{array}
\right)=
\left(
  \begin{array}{c}
    0 \\
    0 \\
  \end{array}
\right)
~~~~
\left(
  \begin{array}{c}
    y^{c2}_1 \\
    y^{c2}_2 \\
  \end{array}
\right)=
\left(
  \begin{array}{c}
    1 \\
    1 \\
  \end{array}
\right)
\end{align}

As no hourly bid of MIC order $c1$ is accepted, its cost is zero, thus the corresponding MIC holds.
The total cost of MIC order $c2$ is 18, while its total income is 24, thus the MIC condition holds.

We can see that assuming this formulation, increasing FT does not work for player 1, as it results in the deactivation of its bid, thus its profit is decreased compared to the reference scenario (Case 1, with truthful bidding). At first glance, it seems that the modification of the objective function resolved the problem of the possibility of strategic bidding.

\subsubsection{Case 3}
\label{Case_3}

In this subsection, we show that if we assume the modified objective function described by eq. (\ref{TSW_2}), which omits the terms corresponding to the hourly bids of MIC orders, and considers the cost of these bids based on FT and VT, another potential problems may arise during the clearing process.

 Let us assume that the true cost of units submitting complex orders is still as
described in subsection \ref{Case_1}. We have seen in subsection \ref{Case_1_mod} that in the case of truthful bidding, the modified objective function (TSW) has no effect on the outcome.

Let us now furthermore assume that player 1 modifies the bid price of the hourly bids belonging to its complex order $c1$ from 1 to 5.5. We assume that all other hourly bids remain unchanged, as summarized in
Table \ref{Table_bids_2}.

\begin{table}[h!]
\begin{center}
\begin{tabular}{|c|c|c|c|c|}
  \hline
  ID & t &q & p & var\\ \hline
  S1 & 1 & 2 & 5 & $y^s_1$\\
  S2 & 1 & 2 & 6 & $y^s_2$\\
  S3 & 2 & 2 & 5 & $y^s_3$\\
  S4 & 2 & 2 & 6 & $y^s_4$\\
  S5 & 1 & 2 & \textbf{5.5} & $y^{c1}_1$\\
  S6 & 2 & 2 & \textbf{5.5} & $y^{c1}_2$\\
  S7 & 1 & 2 & 4 & $y^{c2}_1$\\
  S8 & 2 & 2 & 4 & $y^{c2}_2$\\
  D1 & 1 & -5 & 10 & $y^{d}_1$\\
  D2 & 2 & -5 & 10 & $y^{d}_2$\\
  \hline
\end{tabular}
\end{center}
\caption{Hourly bids of example I: Parameters and corresponding variables. \label{Table_bids_2}}
\end{table}

Furthermore, we suppose that regarding FT and VT, true values are submitted.
\begin{align}\label{MIC_ex_1_mod}
& FT_1=10~~~VT_1=2 \nonumber \\
&  FT_2=10~~~VT_2=2
\end{align}

\begin{figure}[h!]
  \centering
  \includegraphics[width=10cm]{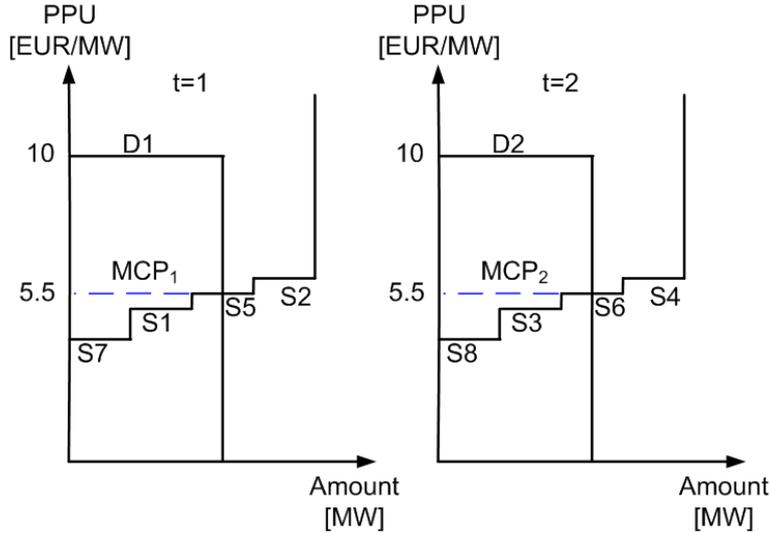}~
  \caption{Resulting dispatch if both MICs would be accepted.}\label{MIC_case_3}
\end{figure}

Let us first note that in this case it is impossible to accept both complex bids.
Fig \ref{MIC_case_3} depicts the resulting dispatch in the case if both MICs are accepted.

As it can be seen, the bid acceptance constraints, which connect the bid prices of hourly bids to the MCP result in the MCP of 5.5 for both periods, implying that the hourly bids of $c1$ are partially accepted.
In this case, the total cost of the MIC order $c1$ is 10+4=14, while its income is 11, thus the MIC condition does not hold. This shows that $c1$ and $c2$ can not be accepted in the same time.
Furthermore, it can be seen in Fig. \ref{MIC_case_3}, that any price for the hourly bid between 5 and 6 will do the job: The implied MCP and thus the income of $c1$ will be different, but even an MCP of 6 and the implied income of 12 will not satisfy the MIC constraint (12<10+4).

Either $c1$ or $c2$ can be accepted, furthermore, as the hourly bids of complex orders are not considered in the modified objective function (the modified TSW), and the parameters $FT$ and $VT$ are the same for $c1$ and $c2$, they result in the same TSW. In both cases, according to eq. (\ref{TSW_2}), the resulting TSW may be calculated as the value of the accepted hourly demand bids (100), minus the value of accepted hourly bids, which are not part of complex orders (S1 and S3 are relevant, with the value of 20), minus the cost of MIC bids according to equation \ref{MIC_cond_basic_accept}, namely 10+4=14, resulting in the TSW value of 66, considering either the acceptance of $c1$ or $c2$. In either case, the resulting dispatch will look like the one depicted in Fig. \ref{MIC_case_2}, resulting in the MCP of 6, and implying an income of 24 for the accepted MIC order (in contrast to the original 20, while the cost is the same).

This means that the objective function has no unique maximum in this case. In such scenarios, the outcome of the market clearing depends on the implementation of the optimization problem, and on the properties of the used solver as well. In general it can be said, that if player 1 modifies its bid parameters as in Table \ref{Table_bids_2}, it is possible that during the clearing $c1$ will be favoured, $c2$ will be 'pushed out' again, and the profit of $c1$ will be increased. More importantly, the resulting optimization problem has no unique solution, and the outcome of the market may depend on implementation details.

\section{Discussion}
\label{sec_discussion}
The modification of the objective function as proposed in eq. (\ref{TSW_2}) may prevent strategic bidding through the manipulation of the submitted FT value, but this modification also implies that strategic bidding trough the bid parameters of hourly bids becomes potentially possible.

 Regarding the principle of the phenomena, the examples presented in the paper demonstrate that if a complex order brings large benefits for the objective function, but it can be accepted only under certain circumstances (e.g. the MCP must be higher than a given value, as in the examples above for $c1$), the solver tends to drop other complex orders with less significant contribution to meet these requirements.

If we consider the original objective function described in eq. (\ref{TSW}), which does not consider the production costs computed from the FT and VT values, the minimum income conditions based on these values may raise issues.

On the other hand, if we neglect the contribution of hourly bids belonging to complex orders, but we do consider the FT/VT-based costs in the objective as in eq. (\ref{TSW_2}), the bid acceptance constraints related to hourly bids can make some dispatches impossible

One may raise the question if the inclusion of both the hourly bid-based components and the FT/VT-based components in the objective function is possible. Theoretically this can be done, however in this case the cost of these supply bids will be considered twice in the objective, which will imply that standard supply bids will be preferred compared to MIC bids during the clearing.

In addition, let us note that irrespective of which formulation (eq. (\ref{TSW_basic}) or eq. (\ref{TSW_2}) ) of the TSW is used, it is possible that players provide false FT and VT values to the ISO. As we have seen, the consequences of this depend on the other bids present in the market and also on the clearing mechanism used, but in any case, such decisions may create the potential of gaming. A possible approach could be to fix these parameters during the registration of the users in the market, and allow their modification only rarely (e.g. once or twice a year). It is also possible that VT is taken into account as the function of the actual fuel prices. This approach may prevent that the submitted FT and VT values are adjusted according to actual market states to maximize the expected income.

Let us furthermore note that the implicit assumption of perfect information has been used through the paper. In realistic cases, the validity of this assumption depends on the publicly available data of DAPXs. The general idea of the paper may be used however also without perfect information. If an MIC order submitted to a DAPX with low hourly bid price values is regularly accepted (thus the contribution to the TSW is large), the bidder may try to increase the submitted FT value in order to test, weather the order is able to 'push out' other MIC bids from the dispatch and increase the resulting MCP.

\section{Conclusions and future work}
\label{sec_conclusions}
In this paper, we have shown how the formalism of minimum income condition orders allows strategic bidding trough the manipulation of various bid parameters and trough the interplay of multiple MIC orders. In addition, we have also shown that various modifications of the objective function used in the market clearing (TSW) only partially resolve this issue.
As these orders are widely used in various DAPXs (mainly because they make the bidding process of generating units easier), and according to the current result they may open possibilities for strategic bidding, further research of the discussed topic is advised.

Additional studies are necessary to determine the practical implications of the discussed theoretical results. The scale of these possible practical implications depends on the typical number and parameters of MIC orders submitted to various markets. Further studies considering real market data (as in \cite{madani2014minimizing} in the case of block orders) can provide results about the practical relevance of the results presented in this paper.

Furthermore, new approaches in the computational formulation of MIC orders and market clearing algorithms may possibly alleviate undesired properties and effects of MIC orders. In particular, innovative formulation of minimum income condition orders, as described in \cite{madani2018revisiting}, where the general class of so called 'MP bids' covers the MIC orders as well, may be free of the disadvantages discussed in this paper -- this, however must be the subject of future studies.

\section*{Acknowledgements}

This work has been supported by the Funds PD 123900 and K 131545 of the Hungarian National Research, Development and Innovation Office, and by the J\'{a}nos Bolyai Research Scholarship of the Hungarian Academy of Sciences.

\section*{Bibliography}
 \bibliographystyle{elsarticle-num}

 \bibliography{energy_GT_1_abr}





\end{document}